\documentstyle[epsfig,12pt]{article}
\begin{document}
\title{Does OPERA probe that the Earth is moving~?}
\author{Dominique Monderen}
\maketitle
\begin{abstract}
The OPERA experiment reported recently a puzzling result. The time of flight of a neutrino beam between the CERN and the Gran Sasso Laboratory has been measured to be slightly shorter than expected. More precisely, an early arrival time of the neutrino with respect to the one computed assuming the speed of light in vacuum of 60.7~ns was measured, with a quite high confidence level. Alternatively, one can conclude that the neutrinos travelled 18.2~m more than light would do in vacuum. In this short paper, we suggest a possible systematic effect that does not appear in the analysis and which can easily been probed to be confirmed.
\end{abstract}
\section*{Introduction}
In order to measure the velocity of a particle, one way consists to measure the distance and the time of flight between the source and the detector, and to compute the ratio to obtain the desired measure. This job has been done with a remarkable care by the OPERA experiment \cite{opera}. The baseline of about 730~km has been measured with an accuracy of 20~cm and the synchronisation of the clocks at CERN and Gran Sasso allows to measure the time of flight with a precision of a few nanoseconds. Computing the ratio, the OPERA experiment obtained a velocity slightly higher than the speed of light, which is now known as the neutrino velocity anomaly. Although the high significance of the result, the effect is small: we speak about a relative effect of $2.48 \times 10^{-5}$. It is therefore still not completely excluded that some tricky systematic effect has been eluded so far. This explains also why the OPERA experiment explicitly refuses any theoretical or phenomenological interpretation of the results.

So far interpretations bloomed as flowers on trees at spring but only a few tried to reconcile the experiment with the consequence of the solid Einstein's postulate: no particle can travel faster than the speed of light. We will here instead investigate the consequences of a very simple observation that can be expressed in a few words: the Earth is moving and thus the target detector is not fixed. This movement has to be included to evaluate the travelling distance.

\section*{The orbital shift}

Because the Earth is moving, the target detector is slightly shifted with respect to the CERN during the time of flight and the true distance experienced by the neutrino beam can be slightly modified. The resulting movement of the target detector induced by the Earth movement can be broken down into two components: the orbital movement and the rotation around the pole axis. The main component is the orbital movement. We will estimate later the other one. It is worth it to notice that this is here a classic Newtonian effect and not a relativistic effect, which is negligible as we will check later.

We will make some approximations in order to evaluate the small shift which affects the results of OPERA. We will consider that the Earth's orbit is circular, with a constant speed. The eccentricity of the orbit is anyway small and can be neglected for our purpose. We will use the following values to evaluate the shift:
\begin{eqnarray*}
\begin{array}{lcl}
{\rm Orbital \, circumference} &=& 924,375,700 \, km
\\
{\rm Orbital \, period } &=& 365.256 \, d = 31,558,118 \, s
\\
{\rm Orbital \, speed } &=& 29,291.22 \, m/s
\end{array} \\
\end{eqnarray*}

We will also use the value of the distance between the CERN and Gran Sasso as measured by OPERA to deduce the distance travelled by the Earth on its orbit during the time of flight of the neutrinos:
\begin{eqnarray*}
\begin{array}{lcl}
{\rm Distance} &=& 730,534 \, m
\\
{\rm Speed \, of \, light} &=& 299,792,458 \, m/s
\\
{\rm Orbital \,shift} &=& 71.38 \, m
\end{array} \\
\end{eqnarray*}

The orbital shift is large enough to provide an explanation of the neutrino velocity anomaly, which is only of $18.2 \, m$. Of course, this shift is not aligned with the trajectory of the neutrino beam and so the consequence on the travelled distance is never maximal. So we need to reconstruct the geometry of the experiment. We will do this in the terrestrial frame of reference. We will use as origin the center of the Earth. The $x$ axis will be given by the direction of the Greenwich meridian; the $y$ axis, by the $90^{\rm o}$ meridian and the $z$ axis, by the pole axis. With the approximations we made, the Earth's trajectory is perpendicular to the Earth-Sun direction. So, when the $x$ axis points to the Sun, {\it i.e.} when it is midday at Greenwich, the shift is aligned with the $y$ axis. To deduce the direction of the shift at any time, it is simple to add a rotation of $15^{\rm o}$ per hour.

\begin{figure}[ht]
 \centering
  \includegraphics[width=12cm]{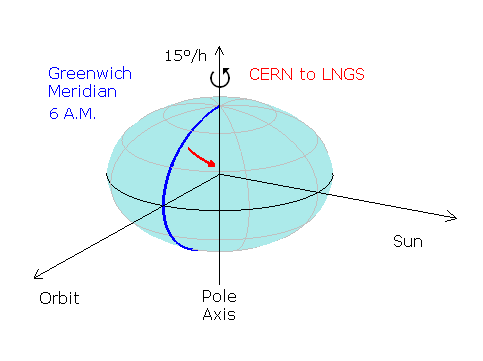}
 \caption{\label{geom} Simplified view without the axial tilt. The position of the Greenwich meridian is illustrated for $\alpha = 0$, {\it i.e.} at 6 AM. The direction of the CERN to LNGS baseline is also shown and its distance has been extended by a factor 3.}
\end{figure}

If the pole axis would be perpendicular to the orbit, what actually happens only twice a year at solstices, the shift would simply be given by:
\begin{eqnarray*}
x_{\rm shift} &=& \cos (\alpha) \, 71.38 \, m
\\
y_{\rm shift} &=& \sin (\alpha) \, 71.38 \, m 
\\
z_{\rm shift} &=& 0 \, m,
\end{eqnarray*}

\noindent where $\alpha$ is the angle between the $x$ axis and the orbit. So for example at midday at Greenwich, the value of $\alpha$ is $-90^{\rm o}$.

This shift is then applied to the coordinates of the target detector in order to compute the distance including the orbital shift. We will use here the following coordinates: the CERN at $46^{\rm o}14' N \, 6^{\rm o}03' E$ and the Gran Sasso Laboratory (LNGS) at $42^{\rm o}25' N \, 13^{\rm o}23' E$. We will also simply assume that those two points are on a perfectly spheric Earth with a radius of $6371 \, km$. The distance between those two points is then $720,513 \, m$, which is sufficiently close to the distance measured by OPERA to be able to illustrate the effect. Of course, one should take more precise coordinates to be accurate. In Table.~\ref{tab:basic_shift}, we illustrate the effect for a few daily values.

\begin{table}[ht]
\centering
\begin{tabular}{|r|r|rrr|rr|}
\hline
\noalign{\smallskip}
\multicolumn{1}{|c|}{Greenwich}&\multicolumn{1}{c}{Angle}
&\multicolumn{1}{|c}{$x_{\rm shift}$}&\multicolumn{1}{c}{$y_{\rm shift}$}&\multicolumn{1}{c}{$z_{\rm shift}$}
&\multicolumn{1}{|c}{Distance}&\multicolumn{1}{c|}{Orbital} \\
\multicolumn{1}{|c|}{Time}& \multicolumn{1}{c}{$\alpha$}
&\multicolumn{1}{|c}{($m$)}&\multicolumn{1}{c}{($m$)}&\multicolumn{1}{c}{($m$)}
&\multicolumn{1}{|c}{($m$)} &\multicolumn{1}{c|}{Shift} \\
\hline
\hline
no shift & & & & & $720,512.56 $ & \\
\hline
12 A.M.& $-90^{\rm o}$ &$0 $ &$-71.38 $ &$0 $ & $720,450.73$ & $-61.84 \, m$ \\
3 P.M. & $-135^{\rm o}$ &$-50,47 $ &$-50,47 $ &$0 $ & $720,455.30$ & $-57.26 \, m$ \\
6 P.M. & $180^{\rm o}$ &$-71,38 $ &$0 $ &$0 $ & $720,493.42$ & $-19.15 \, m$ \\
9 P.M. & $135^{\rm o}$ &$-50,47 $ &$50,47 $ &$0 $ & $720,542.75$ & $30.19 \, m$ \\
12 P.M.& $90^{\rm o}$ &$0 $ &$71.38 $ &$0 $ & $720,574.40$ & $61.84 \, m$ \\
3 A.M. & $45^{\rm o}$ &$50,47 $ &$50,47 $ &$0 $ & $720,569.83$ & $57.27 \, m$ \\
6 A.M. & $0^{\rm o}$ &$71,38 $ &$0 $ &$0 $ & $720,531.71$ & $19.15 \, m$ \\
9 A.M. & $-45^{\rm o}$ &$50,47 $ &$-50,47 $ &$0 $ & $720,482.38$ & $-30.18 \, m$ \\
\hline
{\bf Mean}& & $0 $& $0 $&$0 $ & $720,512.56 $ & $0 \, m$ \\
\hline
\end{tabular}
\caption[]{\baselineskip=12pt The orbital shifts assuming no axial tilt and the corrected travelling distances of the neutrino beam. The mean daily effect is also given.
\label{tab:basic_shift}}
\end{table}

If the neutrino events at LNGS are equally distributed among daytime, the effect fades completely, as the shift at some time is compensated by the opposite shift 12 hours later. The results of the OPERA experiment can be interpreted as a measure of this mean shift. The measure means then that a mean shift of $-18.2 \, m$ is observed. This suggests {\em a priori} that the neutrino events are more frequent during day, which seems a reasonable guess. The obvious way to check whether the orbital shift is able to explain the neutrino velocity anomaly is to check if the time distribution of the events and a rough estimation of the mean orbital shift are compatible with our first conclusion.

\section*{Including the axial tilt}

Some additional care should be taken about the axial tilt of the Earth. Indeed, the pole axis is not perpendicular with the orbital trajectory. For example at autumnal equinox the pole axis is inclined by a maximal positive angle of $23^{\rm o}26'$ with respect to the direction of the Earth movement (the North Pole points forwards). At vernal equinox, on the contrary, the pole axis makes a negative angle (the South Pole points forwards). And therefore the orbital shift should be written as
\begin{eqnarray*}
x_{\rm shift} &=& \sin(\beta) \, \cos (\alpha) \, 71.38 \, m
\\
y_{\rm shift} &=& \sin(\beta) \, \sin (\alpha) \, 71.38 \, m 
\\
z_{\rm shift} &=& \cos(\beta) \, 71.38 \, m
\end{eqnarray*}
\noindent where $\beta$ is the angle between the orbital trajectory and the pole axis. To evaluate it, we can use this relation, valid in the limit of our approximations
\begin{eqnarray*}
\cos(\beta) &=& \cos \left( 2 \pi \frac{{\rm date} - {\rm date}_{AE}}{360.256} \right) \, \sin(23^{\rm o}26') \\
\end{eqnarray*}

\noindent where $({\rm date} - {\rm date}_{AE})$ simply evaluates the number of days since the autumnal equinox. To make it clear, this is simply the projection of the direction of the orbital trajectory on the pole axis in two steps. The first step projects the direction of the trajectory on the projection of the pole axis in the ecliptic plane and the second one projects the result on the pole axis which makes an angle of $(90^{\rm o}-23^{\rm o}26')$ with the ecliptic plane.

We will not fully simulate here the effect on the orbital shift as this work only have sense when applied to the database of the neutrino events. We will however illustrate in Table.~\ref{tab:shift_ae} the effect at autumnal equinox, when the seasonal effect on the shift is maximal. 

\begin{table}[ht]
\centering
\begin{tabular}{|r|r|rrr|rr|}
\hline
\noalign{\smallskip}
\multicolumn{1}{|c|}{Greenwich}&\multicolumn{1}{c}{Angle}
&\multicolumn{1}{|c}{$x_{\rm shift}$}&\multicolumn{1}{c}{$y_{\rm shift}$}&\multicolumn{1}{c}{$z_{\rm shift}$}
&\multicolumn{1}{|c}{Distance}&\multicolumn{1}{c|}{Orbital} \\
\multicolumn{1}{|c|}{Time}& \multicolumn{1}{c}{$\alpha$}
&\multicolumn{1}{|c}{($m$)}&\multicolumn{1}{c}{($m$)}&\multicolumn{1}{c}{($m$)}
&\multicolumn{1}{|c}{($m$)} &\multicolumn{1}{c|}{Shift} \\
\hline
\hline
no shift & & & & & $720,512.56 $ & \\
\hline
12 A.M.& $-90^{\rm o}$ &$0 $ &$-65.49 $ &$28.39$ & $720,443.87$ & $-68.69 \, m$ \\
3 P.M. & $-135^{\rm o}$ &$-46.31 $ &$-46.31 $ &$28.39$ & $720,448.06$ & $-64.50 \, m$ \\
6 P.M. & $180^{\rm o}$ &$-65.49 $ &$0 $ &$28.39$ & $720,483.04$ & $-29.53 \, m$ \\
9 P.M. & $135^{\rm o}$ &$-46.31 $ &$46.31 $ &$28.39$ & $720,528.30$ & $15.74 \, m$ \\
12 P.M.& $90^{\rm o}$ &$0 $ &$65.49 $ &$28.39$ & $720,557.34$ & $44.78 \, m$ \\
3 A.M. & $45^{\rm o}$ &$46.31 $ &$46.31 $ &$28.39$ & $720,553.15$ & $40.59 \, m$ \\
6 A.M. & $0^{\rm o}$ &$65.49 $ &$0 $ &$28.39$ & $720,518.18$ & $5.61 \, m$ \\
9 A.M. & $-45^{\rm o}$ &$46.31 $ &$-46.31 $ &$28.39$ & $720,472.91$ & $-39.65 \, m$ \\
\hline
{\bf Mean}& & $0 $& $0 $&$28.39$ & $720,500.61 $ & $-11.96 \, m$ \\
\hline
\end{tabular}
\caption[]{\baselineskip=12pt Orbital shifts taking into account the axial tilt at autumnal equinox ($\beta = 0$) and the corrected travelling distances of the neutrino beam. The mean daily effect is also given.
\label{tab:shift_ae}}
\end{table}

Interestingly the effect does not fade completely away even if the neutrino events are equally distributed among daytime. There is a net effect on a daily scale. Of course, this effect is compensated 6 months later by its opposite, in such a way that the effect is null if the distribution of the events is flat on both the scales of one day and one year. If all the events would have been observed on autumnal equinox, the results of the OPERA experiment could be interpreted as a measure of the axial tilt. But of course the events were observed at various dates, so that this effect only modifies slightly the previous values and conclusions. We do not expect that this seasonal effect could be extracted yet but for a precise evaluation of the mean orbital shift on the full data set, it should however be included.

\section*{The rotational shift}

For the sake of completeness, we briefly discuss the rotation of the Earth and its consequence on the computation of the orbital shift. We evaluate it with the following values.
\begin{eqnarray*}
\begin{array}{ll}
{\rm Mean \, radius \;\;\; \;\;\;} &= 6371 \, km
\\
{\rm Equatorial \, rotation \, velocity} &= 463.31 \, m/s
\\
{\rm Rotation \, velocity \, at \, a \, 45^{o}\, latitude } &= 327.61 \, m/s
\\
{\rm Rotational \, shift \, at \, equator} &= 1.129 \, m
\\
{\rm Rotational \, shift \, at \, a \, 45^{o}\, latitude } &= 0.798 \, m
\end{array} \\
\end{eqnarray*}

We see that the additional shift due to the rotation of the Earth gives us only a tiny correction of about 1\% of the orbital shift calculated before. It can thus be neglected.

\section*{Lorentz invariance and relativistic effects}

A devil's advocate would point out that there is no need to compute any orbital or rotational shift induced by the movement of the Earth as the measure of the speed of light must be strictly the same in any inertial frame of reference. And so far we know, the movement of the Earth during tiny periods of less than $3$ milliseconds can be considered as linear and of constant speed. So we might wonder why it is necessary to include the orbital shift in the systematic effects of the OPERA measurements.

Actually the answer is quite simple. The OPERA experiment measures separately the time of flight and the distance between the source and the target. And afterwards the velocity of the neutrinos is deduced by computing the ratio but... the distance and the time of flight are measured in two different inertial frames. The distance $x$ is measured in a static frame of reference with respect to the Earth\footnote{Actually it is this frame of reference which should be considered as moving as its accompanies the Earth on its orbital movement but this kind of considerations makes poor sense as only the relative velocity between the two frames of reference matters here.} while the time of flight $t'$ experiences the effects of Earth's motion. The relative velocity between the two frames of reference is given mainly by the Earth's orbital speed that we have seen to be about $30 \, km/s$ or $\beta = 10^{-4}$.

To hold simple expressions, we will consider the case of a neutrino beam travelling parallel to the Earth's orbit. In this case the Lorentzian transformation is written 
\begin{eqnarray*}
ct' &= \gamma \, ct + \beta \gamma \, x &= \gamma (1+\beta)\, ct \approx (1+\beta)\, ct \\
x' &= \beta \gamma \, ct + \gamma \, x &= \gamma (1+\beta)\, x \;\approx (1+\beta)\, x
\end{eqnarray*}

\noindent where as usual $\beta$ is $v/c$ and $\gamma$ is given by $\sqrt{\frac{1}{1-\beta^2}}$. We used the fact that the neutrinos travel at the speed of light, {\it i.e.} $ct = x$ and approximated at first order in $\beta$ as its value is small \footnote{There is here no contradiction is the fact that $ct = x$ and $\beta$ is small. Indeed we do not care here about an observer that would accompany the neutrinos along their journey. {\em That} observer would indeed experience relativistic effects with respect to the terrestrial frame of reference.}. So $\gamma$ is just approximated to 1.

Of course, as $\gamma$ is essentially equal to one, we recover the classic Newtonian relation and justify {\em a posteriori} our previous treatment. Any relativistic corrections would be of the order of $10^{-8}$. The orbital shift that we computed is nothing but the term $\beta ct$.

A na\"ive devil's advocate could finally ask why we do not need to include other shifts as the movement of the solar system in the galaxy or the movement of the galaxy itself. The answer has already be written: only the relative velocity between the two frames of reference matters here and thus the experiment would be blind to any additional motion as Lorentz invariance guarantees it.

\section*{Conclusion}

We investigated the consequences of the Earth motion on the results of the OPERA experiment. We showed that the target detector experiences a small shift of $71 m$, due to the orbital movement of the Earth, during the time of flight of the neutrino beam between the CERN and LNGS. The truly travelled distance is therefore slightly modified and affects the OPERA results as a systematic effect.

While a flat distribution of the events among daytime would not affect the results of OPERA significantly, this shift can induce a net effect if the events are mostly observed during day. A day/night effect could easily be extracted from the data set. If confirmed, we provided the elements in order to evaluate the orbital shift at any time and date and to include it in the analysis of the results. We expect that when this systematic effect will be taken into account, the neutrino velocity anomaly will be solved.

We further investigated a possible seasonal effect, due to the axial tilt, which is showed to be effective but smaller and thus probably undetectable yet. We also investigated whether additional shifts are relevant and discussed the consistency of this treatment with the principle of Lorentz invariance.

A few years ago, the CERN indirectly proved the existence of a satellite in orbit around the Earth with an orbital period of nearly 28 days. It would be amusing that the results of OPERA are today probing that the Earth is moving.

\end{document}